\documentstyle[12pt]{article}
\begin{document}
\begin{center}
{\Large {\bf MODELING OF THE POZZOLANIC REACTION KINETICS BETWEEN LIME AND WASTES OF THE SUGAR INDUSTRY BY ELECTRICAL CONDUCTIVITY MEASUREMENT}}\\

\vspace{1cm}
Ernesto Villar-Coci\~na, Eduardo Valencia-Morales, R. Gonz\'alez-Rodr\'iguez and Jes\'us Hern\'andez-Ru\'iz\\

Physics Department, Central University of Las Villas.\\
Santa Clara 54830, Villaclara, CUBA.\\
(September 2001)
\end{center}

\vspace{1cm}
\begin{abstract}

  In this paper the kinetics of the pozzolanic reaction between lime (calcium hydroxide) and sugar cane straw ash with 20 and 30\% of clay burned at 800 and $1000^0$C is studied. A simple experimental technique where the conductivity is the experimental variable was used. A calibration curve was established for correlating the conductivity with the concentration of calcium hydroxide (CH). We elaborated a mathematical model that allows us to describe the process in kinetic - diffusive or kinetic regimen. The fitting of the model by computerized methods lets us determine the parameters that characterize the process: i.e. the diffusion coefficient and reaction rate constant.  The pozzolanic activity is evaluated according to the obtained values of the reaction rate constant. The results show that sugar cane straw ash (SCSA) has a good pozzolanic activity comparable to that of the rice husk ask (RHA).\ \ \\
\end{abstract}.\\\\\\\\


\newpage
\section{Introduction}

  Pozzolans are incorporated as active additions to the portland cement and concrete due to their capacity for reacting to lime, principally originated during the hydration of portland cement. The result of this reaction is the formation of hydraulic compounds.   
\\
  This pozzolanic reaction modifies some properties of the cement and of the resulting concrete. Although lime pozzolan cements were some of the earliest building materials widely used for all kinds of construction, only recently have lime-pozzolan binders become an interesting alternative for social construction in developing countries. Recent studies \cite{Mart} have shown that wastes from the sugar industry, mainly sugar cane straw ash (SCSA), have pozzolanic activity derived from the high content of amorphous silica in this material.\\ 
  Several methods have focused on the assessment of the pozzolanic activity of a material considering the complexity of its "pozzolanicity", where different mechanisms of the pozzolan-calcium hydroxide (CH) interaction exist and where there is a considerable variation in the nature of materials showing this property.\\ 
 It is known that the development of the pozzolan/CH reaction causes the formation of insoluble products, therefore diminishing the CH concentration in the solution. As a consequence, a decrease of conductivity takes place whose change rate depends on whether the reactivity of the pozzolan is higher or lower.\\
  Conductimetric techniques have been applied for different purposes in the field of cement. A considerable number of papers have been published on the application of electrical conductance methods (and its reciprocal, resistance) in the study of the early hydration of Ordinary Portland Cement. Some of the early works being traced back to the 1930s and extending until the present day \cite{Zhu, Petin, Boast, Hammond, Tamas, Abd, McCarter}.\\
  Raask and Bhaskar \cite{Rassk} however, were the first to design a method for evaluating pozzolanic activity by measuring electrical conductivity. This method allows the measuring of the amount of silica dissolved in a solution of hydrofluoric acid (HF) in which the active material is dispersed. From this, a pozzolanic index is calculated.\\  
  Lux\'an et al. \cite{Luxan} proposed a simple and very fast method for the indirect evaluation of the pozzolanic activity of natural pozzolans. It is based on the electrical conductivity measurement of a calcium hydroxide - pozzolan suspension. \\ 
  Afterwards, Sugita et al. \cite{Sugit, Sugita} evaluated the pozzolanic activity of rice husk ash (RHA) according to Lux\'an`s method. They found a good correlation between noncrystaline silica content in RHA samples and the variation in electrical conductivity of RHA/saturated lime suspensions. \\
  In cement pastes containing a pozzolan, Thashiro et al. \cite{Tashiro} proposed a fast method for evaluating pozzolanic activity by measuring electrical resistance (Only 72 h for testing were required).\\ 
  The above methods for evaluating pozzolanic activity are aimed at the quality control of the behavior of these pozzolanic materials rather than at the quantitative aspect of the lime - pozzolan reaction, (this last being the computing of kinetic coefficients). The knowledge of the kinetic coefficients is a good criterion for evaluating the pozzolanic activity of the materials.\\
  As it is known, lime (or cement)-pozzolan reactions are not yet very well understood and have been the subject of investigation of many researchers. The study of the pozzolanic reaction kinetics is of great scientific interest and of technological importance as well. \\
  From the scientific point of view and for numerous industrial applications, simple but accurate models describing these phenomena are important. The development of such models is also of practical and economical importance and constitutes a topic of scientific interest. \\
The more diffused are the Jander \cite{Jander} and Zhuralev \cite{Zhu} models. Other authors \cite{Bezjak, Alujevic, Kanga, Day, Kondo}have developed mathematical models for describing the pozzolanic reaction kinetics, but they do not always agree with the experimental results.  \\
  In this research the electrical conductivity is recorded experimentally. It can be easily determined and correlated with the $Ca (OH)_2$ concentration. This allows us to follow the kinetics of the reaction of the artificial pozzolans (sugar cane straw ash and rice-husk-ash with 20 and 30\% of clay, burned at 800 and $1000^0$C) without such big experimental requirements. We have worked out a mathematical model that allows us to describe the process in a kinetic-diffusive or kinetic regimen. This model fitted by computerized methods lets us determine process parameters, such as the diffusion coefficient and the reaction rate constant. It allows us to evaluate the pozzolanic activity of these materials in a quick and effective way

\section{Experimental}

\subsection{Materials}
 The pozzolanic ashes were obtained from two kinds of biomasses. The sugar cane straw (SCS) was recollected in the vicinity of the sugar factory "10 de Octubre" in the province of Villaclara, Cuba. The rice husk (RH) was obtained from a rice thresher in the same province.\\
Clay was used as an agglutinant, which can be activated thermally and become a pozzolanic material. This allows us to obtain a Solid Block Fuel (SBF) that can also be used as an alternative energy source.\\
A saturated solution of calcium hydroxide, prepared with deionized water and $Ca(OH)_2$ was used. This was valued with chlorhydric acid. The concentration obtained was 0.040 mole/L.

\subsection{SBF obtainment}
A densification process was carried out to obtain the SBF. This process began with the preparation of the biomass.The RH was used under real conditions as agricultural waste, with approximately 5mm of particle size and the presence of characteristic powders. The SCS was used in a dry natural state, after milling fibers between 4 and 5mm of length were obtained.  20 and 30\% of clay (depending on the total weight of SBF) and a constant amount of biomass (200g) were employed. Water was added to the clay-biomass until an appropriate plasticity degree was reached.\\
Thereafter the SBF was compacted by using the Proctor Standard Test \cite{Norma}which reduces the volume, facilitates the transportation and makes its combustion more efficient. 
Finally, the samples were extracted and subjected to a drying process in the open air. 

\subsection{Pozzolanic materials and experimental procedure}
The combustion process was carried out in a muffle furnace at 800 and $1000^0$C. Ashes were screened and a particle size of 0.15mm was fixed. 
  The chemical composition of ashes is shown in Table 1 and 2.\\
  In the conductivity measurement ash/water solutions were also employed. Its use was to find out the contribution to the conductivity of $Na\:^{+},K^{+},Mg^{2+}$ and other ions present in the solution (although in low concentration).\\
For both, the ash/CH sample and the ash/water, 2,19g for each ash were weighed and then 70ml of CH solution and 70ml of deionized water were added respectively. For measuring the variation of conductivity of the ash/CH solution samples, the contribution to conductivity of ash/water was subtracted.\\
The conductivity measurements were made in a METROHM microconductimeter (Sweden) at $28\:^oC \pm 1\: ^oC$ at different times. The samples were closed tightly to avoid carbonation. To correlate the CH concentration with the conductivity of the CH solution a calibration curve was established (Fig. 1). 

\newpage
\begin{center}
\begin{tabular}{|c|c|c|c|c|}\hline\hline
Compound & \multicolumn{4}{|c|}{Materials(Ashes)} \\ \hline
 & RH+20\%clay & RH+30\%clay & SCS+20\%clay & SCS+30\%clay\\ 
 &RHA1          & RHA2         &SCSA1           &SCSA2 \\ \hline\hline
$SiO_{2}$      & 62.69   & 57.30    & 64.02      & 62.62  \\ \hline
$Al_{2}O_{3}$   & 14.93   & 10.36    &11.29       &13.79  \\ \hline 
$Fe_{2}O_{3}$    &  8.60   & 6.11     &6.59        & 7.65 \\ \hline 
$TiO_{2}$      &  0.86   & 0.60     &0.67        & 0.76  \\ \hline
CaO          &  3.15   & 2.90     &3.65        & 4.41  \\ \hline
MgO          &  3.49   & 2.87     &2.76        & 3.14   \\ \hline
$Na_{2}O$      &  2.58   & 1.92     &1.88        & 2.22   \\ \hline
$K_{2}O$       &  1,36   & 1.64     &2.35        & 1.91   \\ \hline
Ignition loss&  2.04   & 14.21    &4.88        & 3.30   \\ \hline
Total        &  99.70  & 97.91    &98.09       &99.80   \\ \hline       

\end{tabular}
\normalsize
\end{center}
{\bf Table 1}.Chemical composition of the materials calcined at 800$^{0}$ C.\\

\begin{center}
\begin{tabular}{|c|c|c|}\hline\hline
Compound & \multicolumn{2}{|c|}{Materials(Ashes)} \\ \hline
 & SCS+20\%clay & SCS+30\%clay \\
 &  SCSA3        & SCSA4         \\ \hline\hline
$SiO_{2}$      & 66.96   & 65.50  \\ \hline
$Al_{2}O_{3}$   & 12.72   & 13.80  \\ \hline 
$Fe_{2}O_{3}$    &  6.94   & 7.65 \\ \hline 
$TiO_{2}$      &  0.71   & 0.76  \\ \hline
CaO          &  4.53   & 3.65    \\ \hline
MgO          &  3.14   & 3.35    \\ \hline
$Na_{2}O$      &  2.02   & 2.22  \\ \hline
$K_{2}O$       &  1.91   & 1.84  \\ \hline
Ignition loss&  0.82   & 0.55    \\ \hline
Total        &  99.75  & 99.32   \\ \hline       

\end{tabular}
\normalsize
\end{center}
{\bf Table 2}.Chemical composition of the SCSA calcined at 1000$^{0}$ C.\\

\section{Formulation of the problem }

 Several methods have been applied to evaluate the pozzolans \cite{McCarter, Rassk, Luxan, Tashiro, Moreau, Sharma, Frias, Cabrera, Talero}. These contemplate chemical, physical and mechanical viewpoints and yield qualitative and quantitative evaluations. They are based, in most cases, on the analysis of the pozzolanic reaction in CH/pozzolan systems. Since neither the mechanism nor the kinetics of the pozzolanic reactions are well understood at present, disparity of opinions about how to evaluate the properties of a pozzolan arise. The pozzolanic reaction is heterogeneous. This can be considered about solid-solution type:\\

\begin{equation}
A_{(L)}+b B_{(S)}\rightarrow E_{(L)}+F_{(S)}
\end{equation}

Some models \cite{Smith} have been proposed for describing the reaction (1). We use the Decreasing Nucleus Model (DNM). In accordance with this model, when solution A (with $C_b$ concentration) reacts on the surface of the B solid reactant (with $r_s$ radius) a layer of reaction products F is formed and the nucleus without reacting (with $r_c$ radius) decreases gradually. If this layer is porous the reaction takes place by diffusion of A through the layer F until it is verified on the interface between F and the nucleus without reacting. The temperature is considered uniform in all the heterogeneous region.\\
  The following suppositions are made:\\
- The spherical form of the granule is kept and the densities of F and B are the same. Consequently, the total radius of the granule $r_s$ doesn't change with time and an intermediate region doesn't exist between the nucleus and the layer of product \cite{Yagi}. \\
- The movement rate of the reaction interface $dr_C/dt$ is small in comparison to the diffusion speed of A through the product layer (pseudo-stable state) \cite{Bischoff}. This is valid when the density of the fluid in the pores of F is smaller than the density of the solid reactant, which is certain in general.\\
In fig.2 a scheme of the concentration profiles according to DNM is shown. \\
In the case of pozzolan/lime solution system the reaction occurs through the following stages \cite{Rabilero}:\\

$\bullet$  Diffusion of $Ca\:^{2+}$ ions from the solution to the surface of the pozzolan particle.\\
$\bullet$  Adsorption of $Ca\:^{2+}$ ions on the surface of the pozzolan particle.\\
$\bullet$  Diffusion of $Ca\:^{2+}$ ions through the layer of reaction products.\\
$\bullet$  Chemical interaction between $Ca\:^{2+}$ ions and the pozzolan particle.\\

Under pseudo-stable state conditions, the speed equations, expressed as mole of solution A (CH solution) that disappears per unit of time per particle are identical:\\

\begin{equation}
\frac{d N_A}{d t}=4\pi r_{s}^2 K_m [(C_A)_b - (C_A)_S] \:\:\:\:\mbox{external diffusion}\
\end{equation}

\begin{equation}
\frac{d N_A}{d t}=4\pi r_{s}^2 D_e [\frac{d C_A}{d t}]_{r=r_c} \:\:\mbox{diffusion through the product}\
\end{equation}

\begin{equation}
\frac{d N_A}{d t}=4\pi r_{c}^2 D_e (C_A)_c \:\:\mbox{reaction on the interface} \:r=r_c\
\end{equation}

In eq. 4 a $1^{st}$ order irreversible chemical reaction is assumed.\\
If we handle the equation (2), (3) and (4), we obtain one that relates the decreasing of the nucleus in terms of the exterior concentration of CH with time.\\

\begin{equation}
\frac{-d r_c}{d t}=\frac{b m_B K (C_A)_b}{\varrho_B [1+(\frac{r_c^2}{r_s^2})(\frac{K}{K_m})+(\frac{K\:r_c}{D_e})(1-\frac{r_c}{r_s})]}
\end{equation}

where $K_m$ is the external mass transfer coefficient,$D_e$ is the effective diffusion coefficient of A through of the porous layer of product F and K the reaction rate constant.\\

Since in our case the functional dependence of $r_c$ with time isn't known, it is difficult to apply (5) to the experimental results, that are the variation of  $(C_A)_b$ with t. For that reason it is necessary to postulate a behavior that allows us to relate the external concentration $(C_A)_b$ to time.\\
According to the small particle size and the morphology of the product layer F \cite{Mart} the decrease of the nucleus to react should follow a similar behavior to the change of the concentration (or conductivity) in the outer solution (Fig.3 and 4). This change shows a fast decrease at early ages and an asymptotic behavior for long times. In accordance to the above-mentioned, this behavior of $r_c$ with t is adjusted to a dependence of the type:\\ 

  $r_c = r_s Exp(-n t)$\\\\

where n is related to the diminishing speed of the nucleus.

 Therefore eq.(5) is:\\

\begin{equation}
nr_s Exp(-n t)=\frac{C_t K \frac{4}{3} \pi r_s^3}{ [1+(\frac{KExp(-2nt)K}{K_m})+(\frac{Exp(-nt)\:(1-Exp(-nt)) K\:r_s}{D_e})]}
\end{equation}

where $\frac{b\: m_b}{\varrho_B}= V = \frac{4}{3} \pi r_s^3$ is the volume of the particle.\\
Since the displacement time of the ions in the solution is much smaller that the necessary time for the diffusion and reaction inside the particle, we can consider the external mass transfer coefficient as very large and eq. (6) is reduced to:\\  

\begin{equation}
C_t=\frac{1.59153 Exp(-3nt) [-1+Exp(nt)]n}{D_e}+\frac{10.61032 Exp(-nt)n}{K}
\end{equation}
where $r_s$=0.15mm.\\
Eq. (7) represents a kinetic-diffusive model. This expresses the variation of CH concentration with time. The lineal relationship between the CH concentration and conductivity allows their direct correlation.\\
The pozzolanic reaction, as was previously expressed, consists of different stages. It is known that the resistances of these stages are usually very different and that the stage presenting a bigger resistance (i.e, that lapses more slowly) controls the process. According to this, it is possible in certain cases to have a diffusive (described by the first term of (7)), kinetic (second term) or kinetic-diffusive (both terms) behavior. The fitting of the model (eq.(7)) allows to determine the kinetics coefficients (diffusion coefficient and reaction rate constant) and therefore a rigorous characterization of the process.  \\

\section{Results and Discussion}
  In figures 3 and 4 the conductivity variations versus time for the CH-pozzolans suspensions are shown. A decrease of the electrical conductivity of the suspension is appreciated. This behavior is attributed to the formation of insoluble products with the corresponding decrease of the CH concentration in the solution. As result of this, conductivity decreases.\\
A considerable variation (loss) of conductivity in early ages is appreciated. The stabilization of the curve (to very close to zero values) is reached for big time instants and it depends on the analyzed sample. This indicates when the reaction has finished practically.\\
For pozzolanic samples calcined at $800^0$C a greater reactivity is qualitatively appreciated for SCS1 followed by SCS2, RH1 and RH2 respectively. For $1000^0$C SCS3 is more reactive than SCS4.\\
For comparing data, it is convenient to calculate the loss as relative according to the following equation:\\

\begin{equation}
\xi=\frac{C_0-C_t}{C_0}
\end{equation}

where $\xi$ is the relative loss of conductivity, $C_0$ the initial conductivity and $C_t$ the absolute loss of conductivity with time for pozzolan/lime suspension.\\
Lux\'an et. al.\cite{Luxan} proposed a methodology for indirectly determining the pozzolanic activity of natural pozzolans. They establish an index given by the variation between the initial and final conductivity for a time of only 120s.\\
Recently, Payá et. al. \cite{Paya} proposed a methodology for evaluating flay ash. They calculated the pozzolanic activity as the percentage of loss in conductivity at several reaction times (100, 1000 and 10000s).\\
We propose a kinetic-diffusive model (eq. (7)). The use of eq.(8) allows us to correlate concentration and conductivity taking into account their linear dependence. It also let us work with a dimensionless magnitude.\\
According to eq.(8),  eq.(7) is transformed into:\\

\begin{equation}
\xi=1-(\frac{1.59153}{C_0}\frac{Exp(-3nt) [-1+Exp(nt)]n}{D_e}+
\frac{10.61032}{C_0}\frac{Exp(-nt)n}{K}
\end{equation}

Shown in figures 5, 6, 7, 8, 9 and 10 are the relative loss of conductivity $\xi$ versus time for the CH-pozzolans samples calcined at 800 and $1000^0$C. Solid lines represent the curves of the fitted model. The fitting of the model (eq.(9)) permitted us to determine the parameters n, De and/or K in each case.\\
 In the case of the samples calcined at $800^0$C a kinetic behavior was appreciated.  For this statement was took into account the accuracy in the fitting of the model and statistical parameters such as: correlation coefficient (r), coefficient of multiple determination ($R^2$), 95\% confidence intervals, residual scatter, residual probability and variance analysis. This means that the chemical interaction speed on the surface of the nucleus without reacting is smaller than the diffusion speed of the reactant through the reaction product layer. As a result of this, the general speed of the whole process is determined by the lowest stage (chemical reaction). The values of the n parameter and the reaction rate constant K are given in Table 3. In the figures the correlation and multiple determination coefficients r and $R^2$ are shown.\\

\begin{center}
\begin{tabular}{|c|c|c|}\hline\hline
 Material & n  & Rate reaction constant\\ 
   (Ash)&  & K($h^{-1}$)     \\   \hline\hline
 RHA1   & $1,95.10^{-2}\pm 0,21.10^{-2}$ &$3,34.10^{-2}\pm 0,39.10^{-2}$\\ \hline 
 RHA2   &$1,25.10^{-2}\pm 0,10.10^{-2}$ &$1,98.10^{-2}\pm 0,15.10^{-2}$ \\ \hline
 SCSA1  &$4,72.10^{-2}\pm 0,45.10^{-2}$ &$7,12.10^{-2}\pm 0,52.10^{-2}$\\ \hline
 SCSA2  &$2,40.10^{-2}\pm 0,22.10^{-2}$ &$3,84.10^{-2}\pm 0,32.10^{-2}$\\ \hline

\end{tabular}
\normalsize
\end{center}
{\bf Table 3}. Reaction rate constants and n parameter for pozzolans calcined  at $800^{0}$C.\\

According to these results, the SCS1 sample shows the highest reactivity (bigger K) followed by SCS2, RHA1 and RHA2. These results agree with the qualitative analysis carried out previously. The K value directly reflects the reactivity of the pozzolan and is a direct index of the pozzolanic activity.\\
 In the case of samples calcined at $1000^0$C a kinetic-diffusive behavior is accepted taking into account the same previous considerations. This means that the chemical reaction speed and diffusion speed are comparable under these conditions. Therefore, both processes determine the general speed of the whole process. The values of n, diffusion coefficient ($D_e$) and the reaction rate constant (K)are shown in Table 4.\\

\begin{center}
\begin{tabular}{|c|c|c|c|}\hline\hline
 Material   & n  &   Diffusion coefficient &  Rate reaction constant\\ 
   (Ash)&  &    $D_e(mm^{2}/h)$      &K($h^{-1}$)   \\   \hline\hline
 SCSA3     & $1,43.10^{-2}\pm 0,04.10^{-2}$&$2,48.10^{-3}\pm 0,22.10^{-3}$ &$1,94.10^{-2}\pm 0,07.10^{-2}$\\ \hline 
 SCSA4  &$7,95.10^{-3}\pm 0.22.10^{-3}$&$1,39.10^{-3}\pm 0,12.10^{-3}$ &$1,03.10^{-2}\pm 0,03.10^{-2}$ \\ \hline

\end{tabular}
\normalsize
\end{center}
{\bf Table 4}. Reaction rate constants, diffusion coefficients and n parameter for SCSA pozzolans calcined at $1000^{0}$C.\\

According to these results, the highest reactivity was shown by SCS3 followed by SCS4.\\
 It is appreciated (qualitative and quantitatively) that the SCS ashes diminish their reactivity as the calcination temperature increases. This may be due to the loss of reactivity of SCS silica, due to calcination at a higher temperature that increases the crystallinity of the silica produced.\\
According to the kinetic coefficients K obtained, the SCSA calcined at $800^0$C shows a high reactivity, even above RHA considered of good pozzolanic qualities, guaranteed in numerous works.\\
These results coincide with the results of the experimental test \cite{Mart} (X-rays diffraction (XRD), thermogravimetry (TG), mercury intrusion porosimetry (MIP) and scanning electronic microscopy (SEM), for the pozzolanic characterization of this material.\\

\newpage
\section{Conclusions}

1. The conductometric changes allow the characterization of SCSA as a material with good pozzolanic activity comparable to that of RHA based on the determination of kinetic parameters.\\
2. The kinetic-diffusive model work out permits to the description of the pozzolanic reaction kinetics in CH/(SCSA, RHA) systems, determining the kinetics coefficient (diffusion coefficients and reaction rate constants) in the fitting model process with accuracy. The reaction rate constants give a very exact index of the reactivity or pozzolanic activity of the materials analyzed.\\
3. This method can be used for determining the pozzolanic activity in a direct, economical and rigorous way.\\
4. The ashes obtained starting from burning the SBF show both high reactivity and good pozzolanic qualities. Besides, they may constitutes an alternative source of energy in countries lacking energy reserves that possess high volume of wastes due to their agricultural economy.\\

\section{Acknowledgments}

 The authors wish to thank to Dr. José Fernando Martirena its collaboration.

\newpage

\begin{Large}
Figure Captions
\end{Large}

{\bf Figure 1}. Conductivity-concentration calibration curve.\\\\\\

{\bf Figure 2}. Scheme showing concentration variations according to DNM.  \\\\\\

{\bf Figure 3}. Variation of conductivity with time for samples burned at $800^0$C.\\\\\\

{\bf Figure 4}. Variation of conductivity with time for samples burned at $1000^0$C. \\\\\\

{\bf Figure 5}. Relative loss in conductivity for RHA1 burned at $800^0$C.\\
               $\bullet$  Experimental $\;\;--$ Model\\\\\\

{\bf Figure 6}. Relative loss in conductivity for RHA2 burned at $800^0$C.\\
               $\bullet$  Experimental $\;\;--$ Model\\\\\\

{\bf Figure 7}. Relative loss in conductivity for SCSA1 burned at $800^0$C.\\
               $\bullet$  Experimental $\;\;--$ Model\\\\\\

{\bf Figure 8}. Relative loss in conductivity for SCSA2 burned at $800^0$C.\\
               $\bullet$  Experimental $\;\;--$ Model\\\\\\

{\bf Figure 9}. Relative loss in conductivity for SCSA3 burned at $1000^0$C.\\
               $\bullet$  Experimental $\;\;--$ Model\\\\\\

{\bf Figure 10}. Relative loss in conductivity for SCSA4 burned at $1000^0$C.\\
               $\bullet$  Experimental $\;\;--$ Model\\\\\\

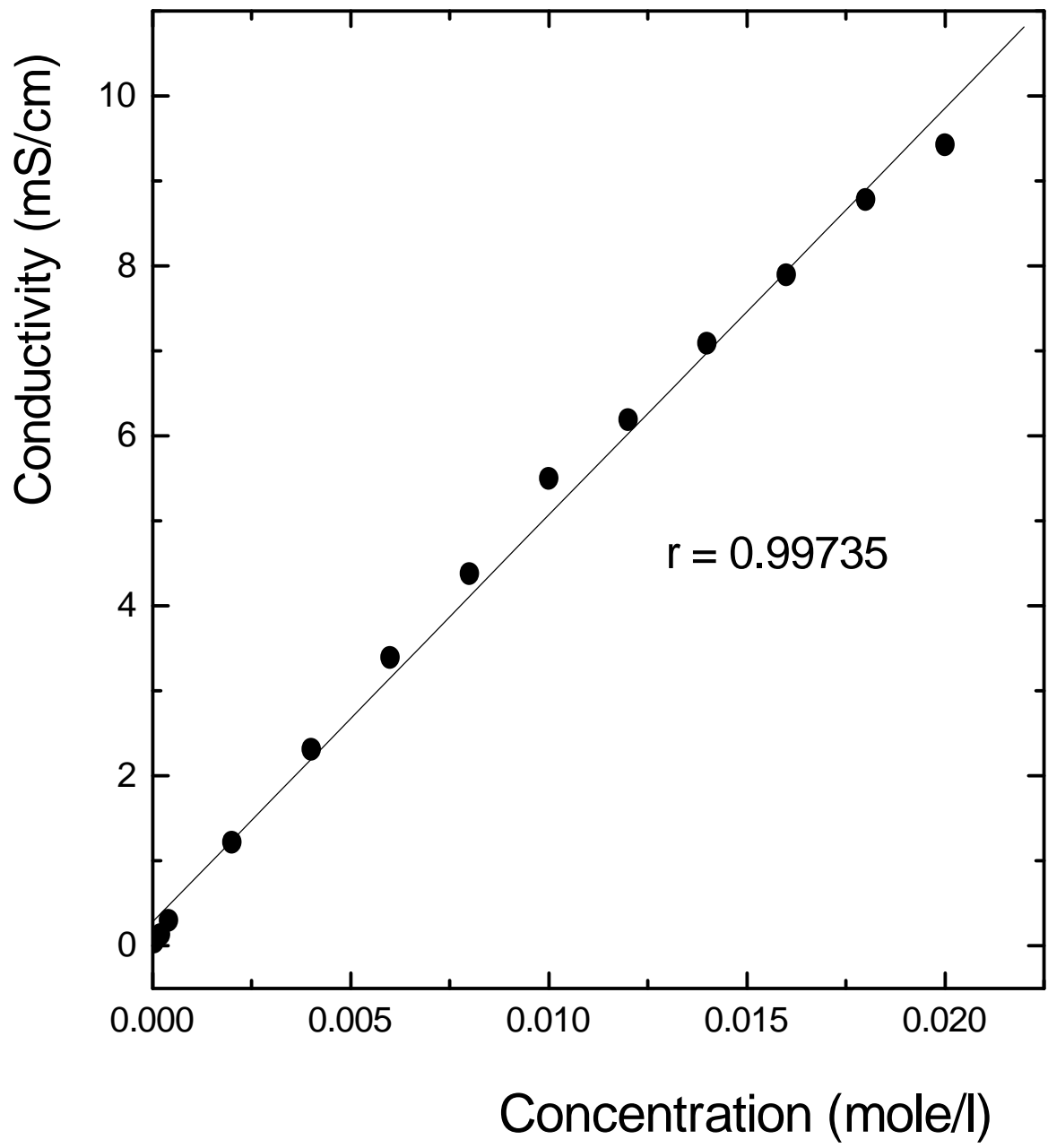
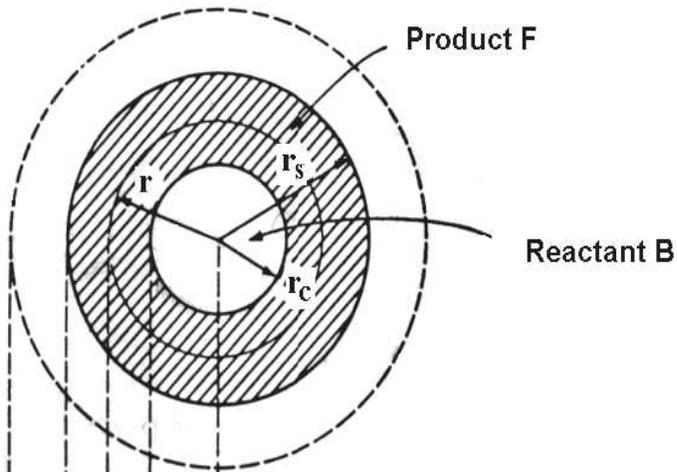
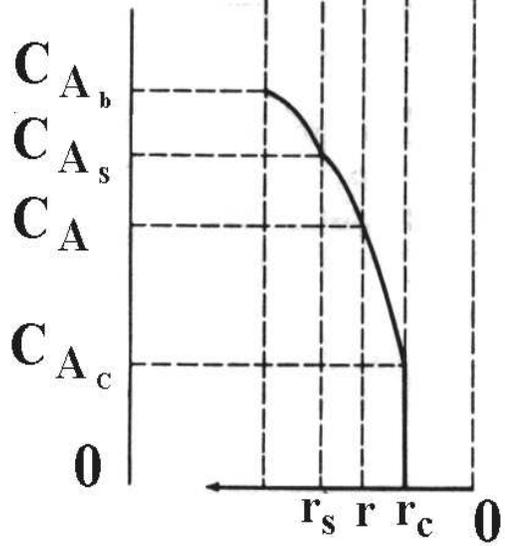
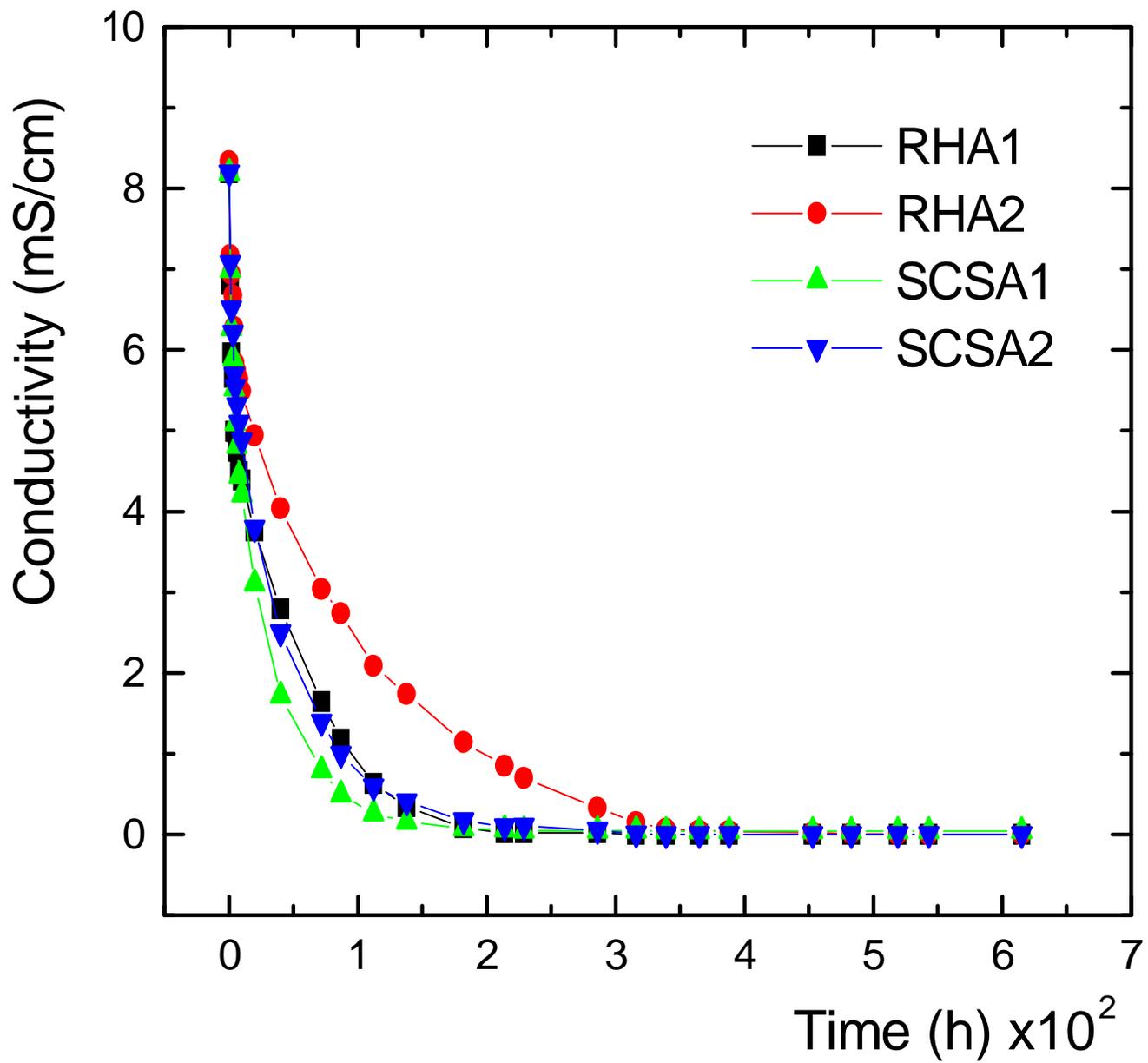
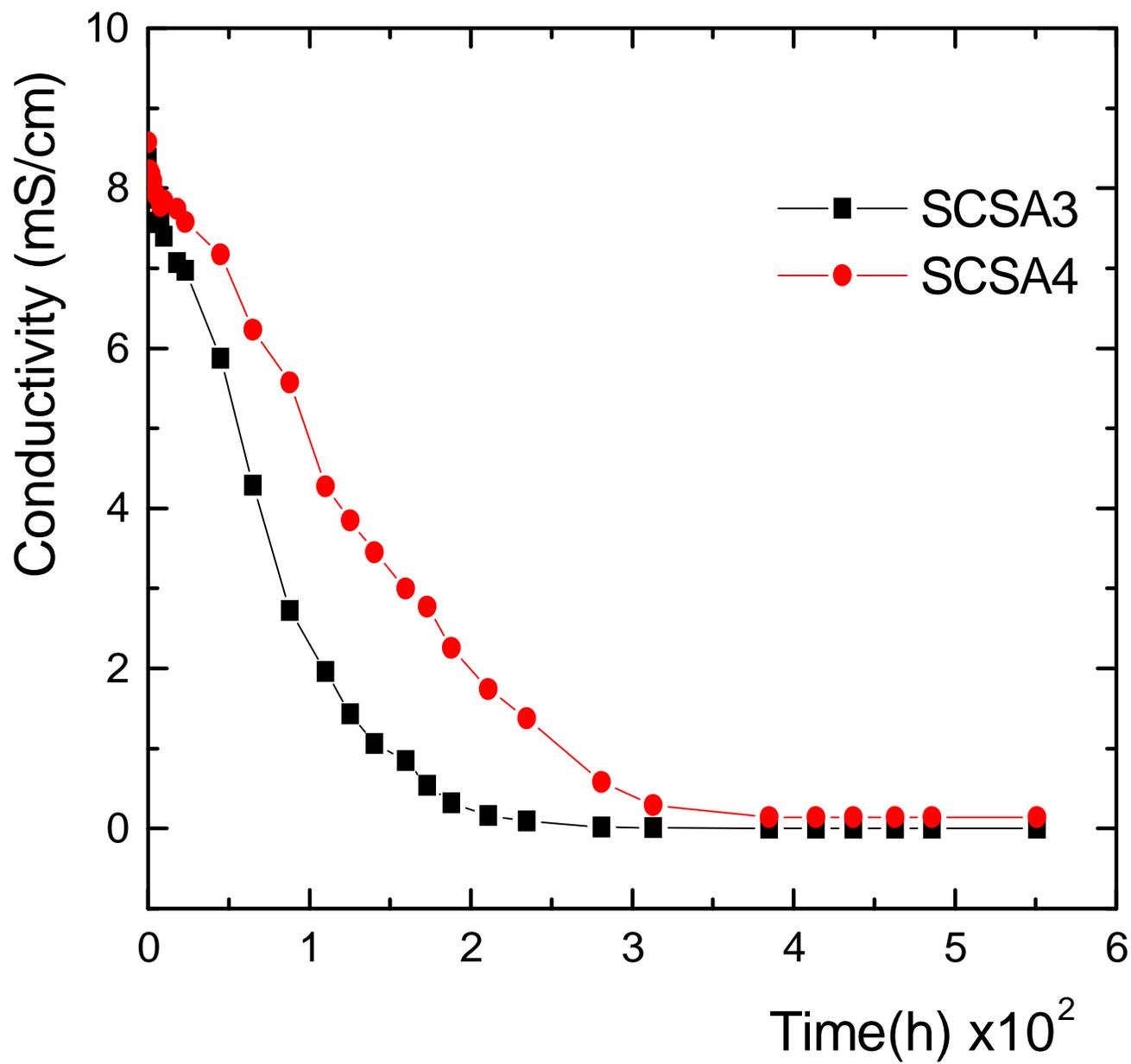
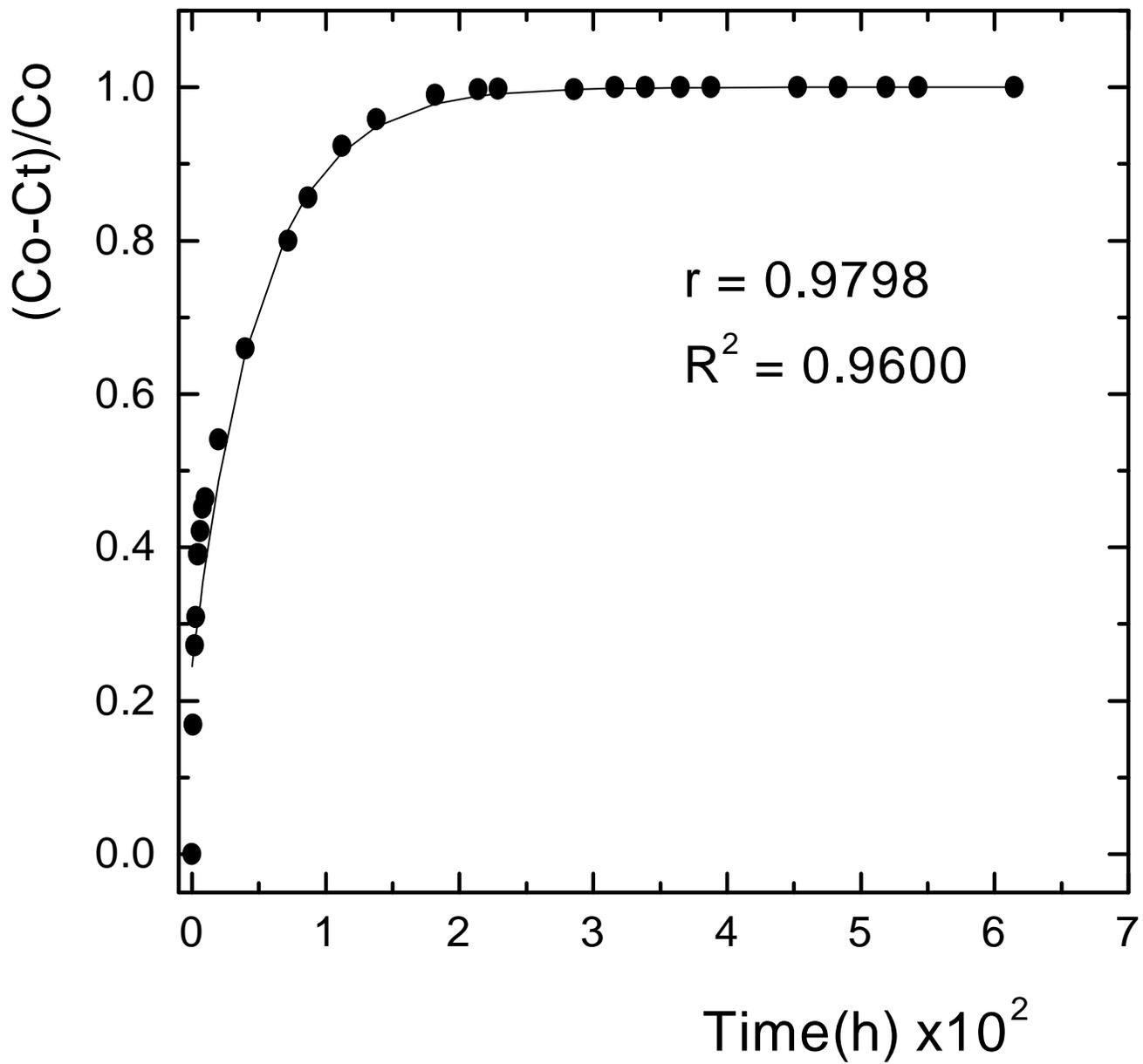
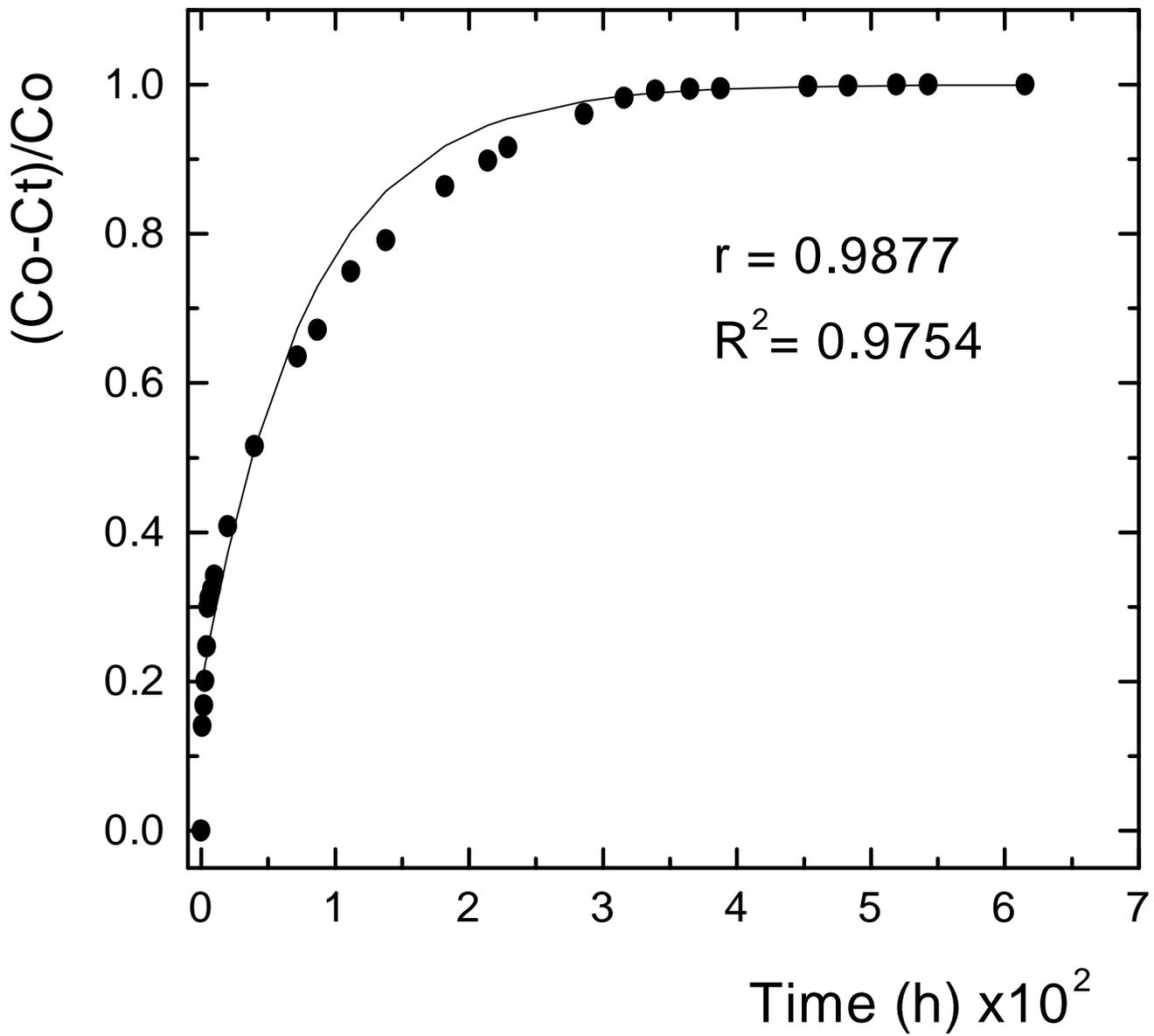
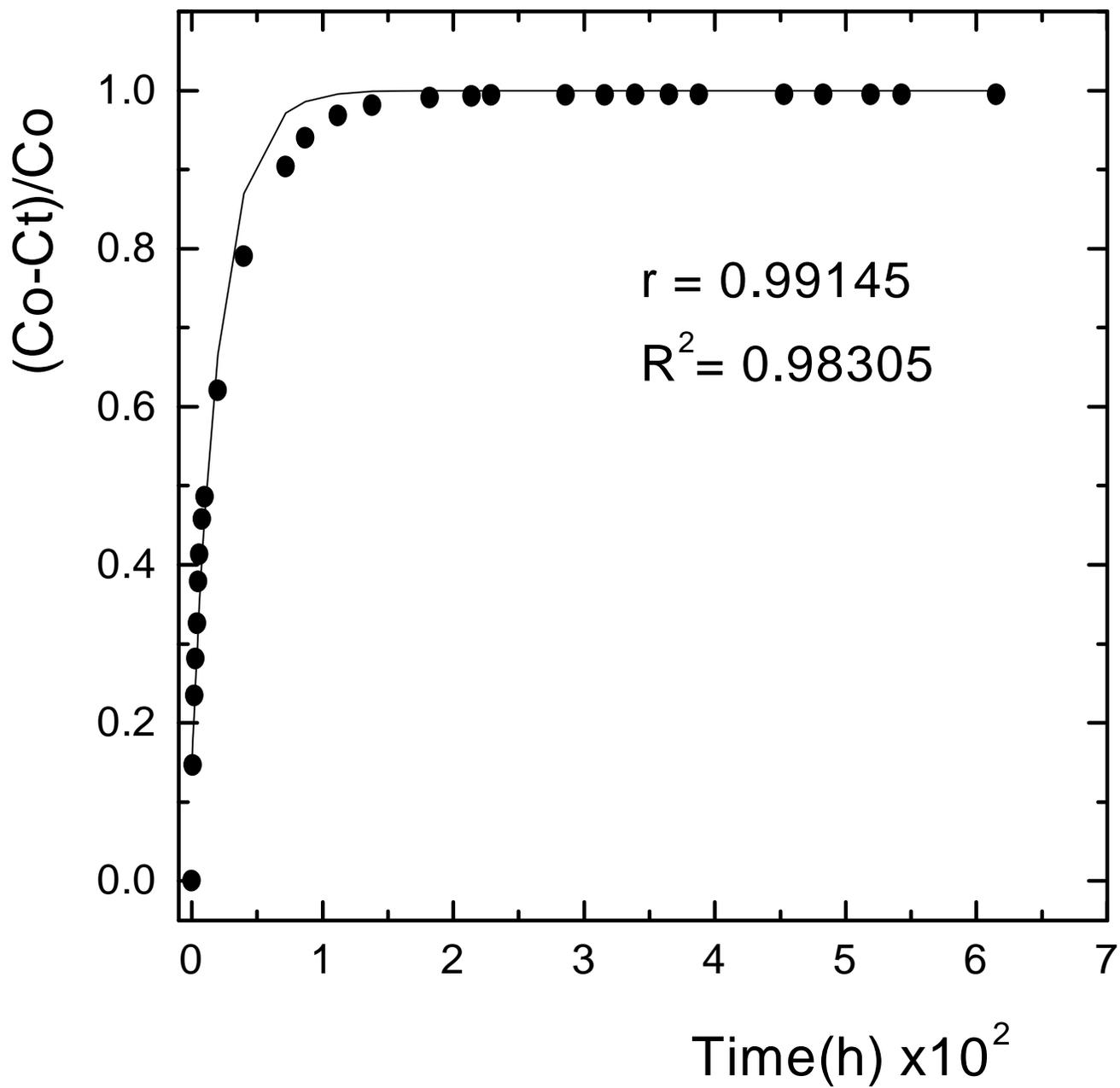
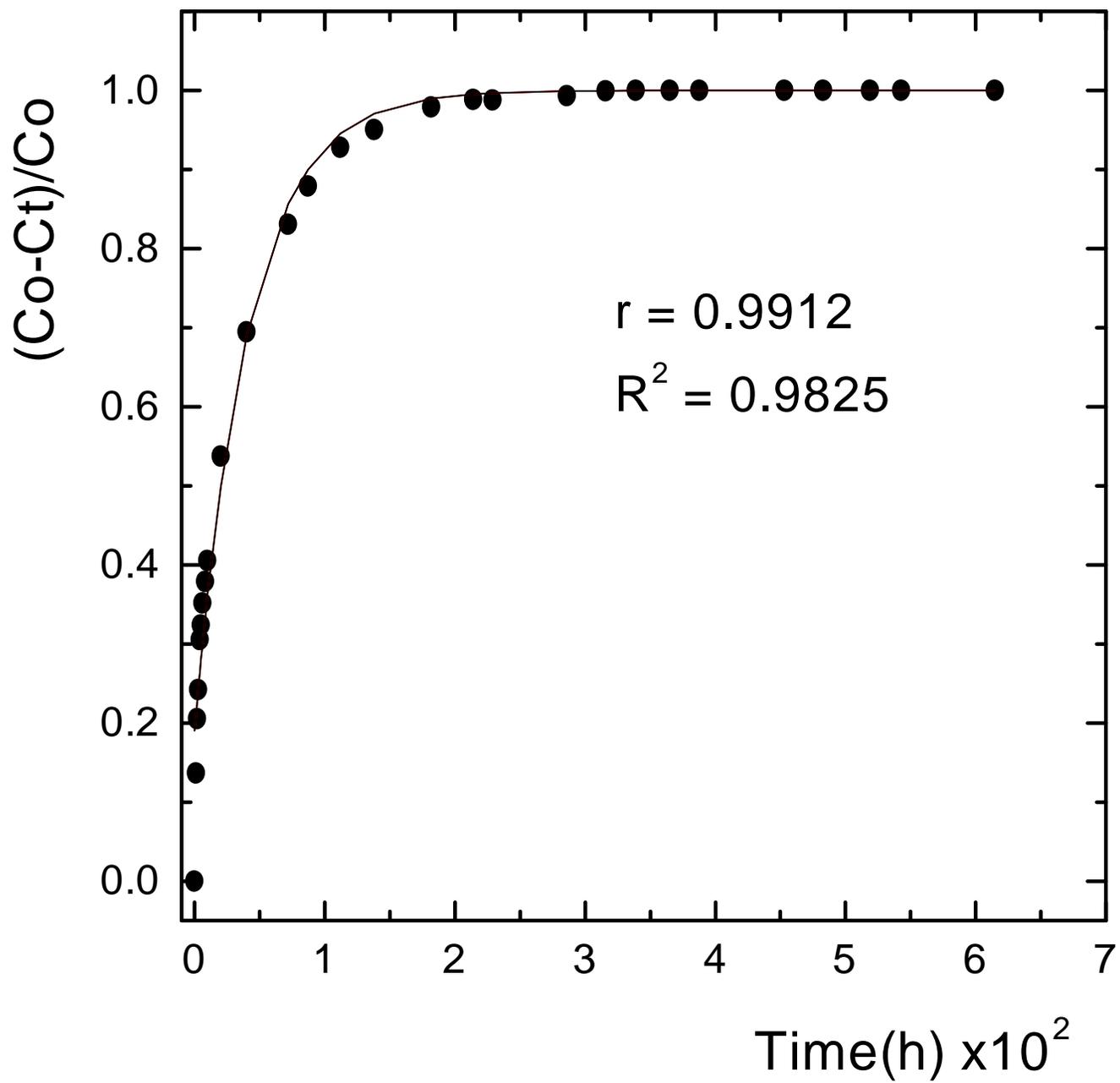
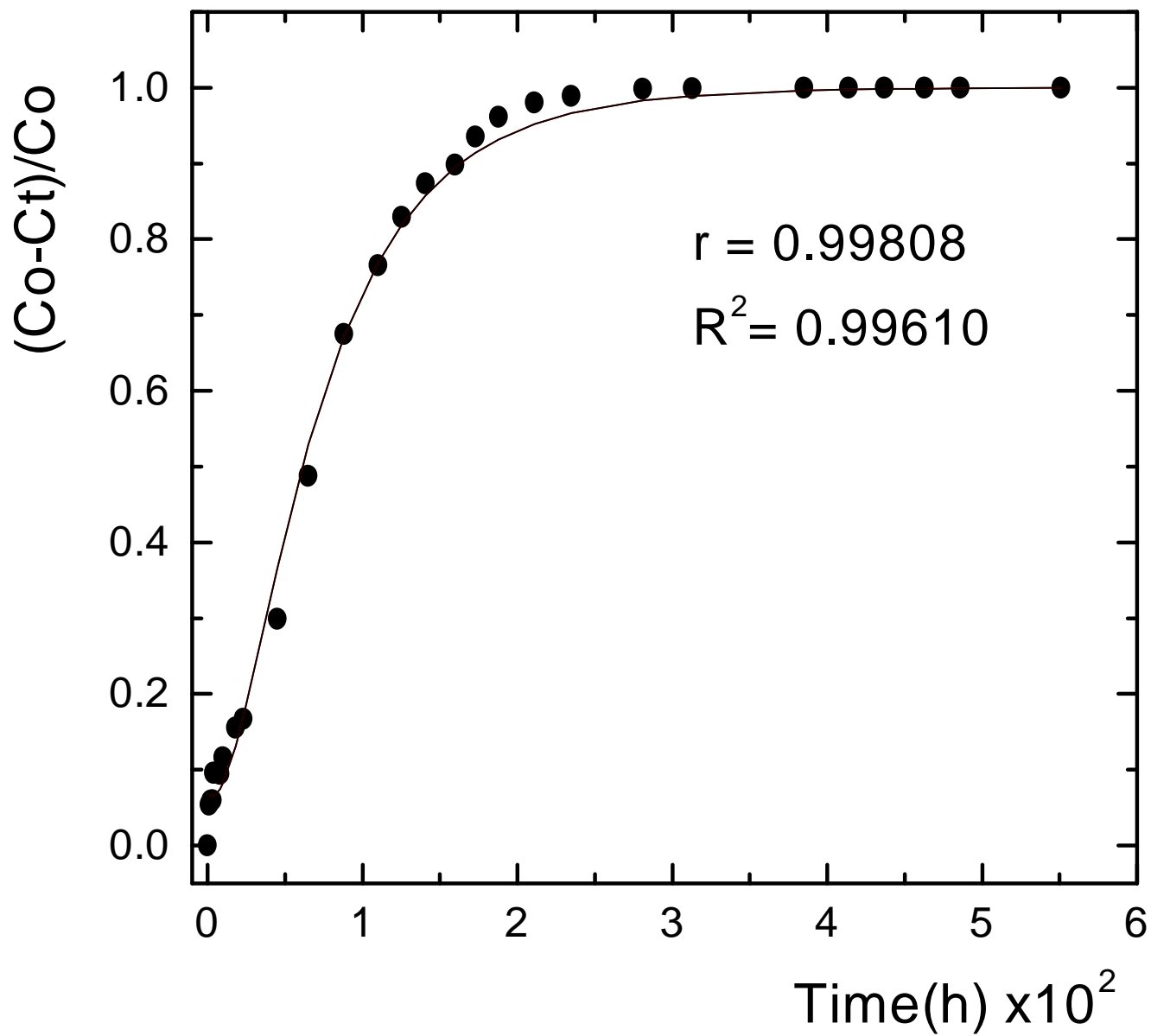
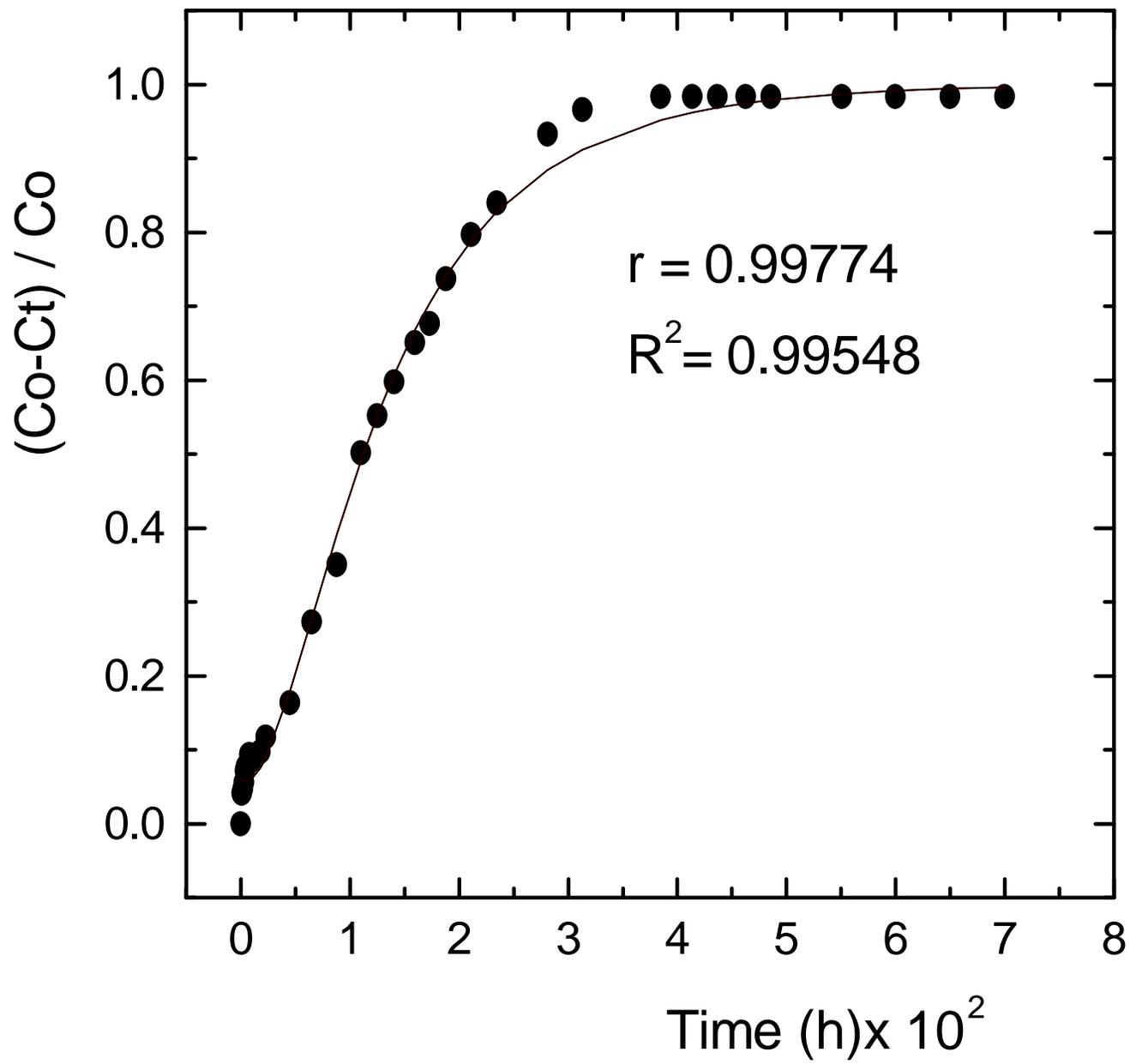
\end{document}